\documentclass[10pt]{article}
\usepackage[english]{babel}
\usepackage{amsmath,amsfonts,amssymb,bm}\interdisplaylinepenalty=2500
\ifx\pdfoutput\undefined
  \usepackage{graphicx,color}   
  \usepackage[nativepdf,backref,bookmarks]{hyperref} 
  \usepackage{rotating}
\else
  \usepackage[backref,bookmarks]{hyperref}  
  \usepackage[pdftex]{graphicx,color}     
  \usepackage[pdftex]{rotating}              
\fi
\graphicspath{{figs-eps/}}
\mathchardef\varphi="011E        \mathchardef\phi="0127
\ifx\undefined\degrees\def\degrees{\ensuremath{^{\circ}}}\fi
\ifx\undefined\celsius\def\celsius{\ensuremath{^{\circ}\mathrm{C}}}\fi
\ifx\undefined\unit\def\unit#1{\ensuremath{\mathrm{\,#1}}}\fi
\ifx\undefined\micro\def\micro{\ensuremath{\mu}}\fi
\ifx\undefined\sups\def\sups#1{\ensuremath{^{\mathrm{#1}}}}\fi
\ifx\undefined\subs\def\subs#1{\ensuremath{_{\mathrm{#1}}}}\fi
\ifx\undefined\ohm\def\ohm{\ensuremath{\mathrm{\Omega}}}\fi

\begin{document}
\title{Flicker noise in high-speed p-i-n photodiodes}
\author{Enrico~Rubiola\thanks{%
	E. Rubiola is with the Universit\'e Henri Poincar\'e, ESSTIN and LPMIA, Nancy, France},\\
	Ertan Salik, Nan Yu, and Lute Maleki\thanks{%
	E. Salik, N. Yu, and L. Maleki are with the Jet Propulsion Laboratory, California Institute of Technology, Pasadena, CA, USA}}

\date{March 2, 2005}

\maketitle

\begin{abstract}
The microwave signal at the output of a photodiode that detects a 
modulated optical beam contains the phase noise $\phi(t)$ and the 
amplitude noise $\alpha(t)$ of the detector.   
Beside the white noise, which is well understood, the spectral
densities $S_\phi(f)$ and $S_\alpha(f)$ show flicker noise,  
proportional to $1/f$. 
We report on the measurement of the phase and amplitude noise 
of high-speed \emph{p-i-n} photodiodes.  
The main result is that the flicker coefficient of the samples is  
$\sim10^{-12}$ \unit{rad^2/Hz} ($-120$ \unit{dBrad^2/Hz}) for phase noise, and
$\sim10^{-12}$ Hz$^{-1}$ ($-120$ dB) for amplitude noise.
These values could be observed only after solving a 
number of experimental problems and in a protected environment.  
By contrast, in ordinary conditions insufficient EMI isolation, and also 
insufficient mechanical isolation, are responsible for additional noise to be taken in.  This suggests that if package and EMC are revisited, applications
can take the full benefit from the surprisingly low noise of the 
\emph{p-i-n} photodiodes.
\end{abstract}


\maketitle

\section{Introduction}\label{sec:introduction}
Many high performance applications of microwave photonics and optics are impacted by phase noise of the microwave signals modulated as sidebands on the optical carrier. Examples of such applications include the frequency distribution system in the NASA
Deep Space Network~\cite{calhoun00ptti}, very long baseline radio astronomy
interferometry arrays (VLBI)~\cite{sato00im}, laboratory
time and frequency comparisons~\cite{narbonneau03fcs,bibey99mtt}, photonic
oscillators \cite{yao96josab,yao97ol}, and laser
metrology \cite{scott01jstqe,ivanov03jstqe}.  The contributions of nearly all microwave and photonic circuit elements to the phase noise is, for most part, well understood, or at least determined experimentally.  This is not the case for the contributions of the photodetector to the phase noise. Many high performance systems such as those mentioned above could be limited by the close-in noise of the photodetector.
Yet the lack of information regarding this topic---only one conference article \cite{maleki98ofc} is found in the literature---made this work necessary. 
In this paper we describe a sensitive measurement technique for the close-in phase noise and  amplitude noise, and the measurement of several photodetectors used to detect microwave (10 GHz) sidebands of optical carriers.

When a light beam is modulated in intensity by a microwave signal and
fed into a photodetector, the detector delivers a copy of the
microwave signal at its output, with added noise.  Flicker noise
is the random fluctuations of the microwave phase and of the fractional amplitude, $\phi(t)$ and $\alpha(t)$, with power spectrum density $S(f)$
proportional to $1/f$.  This refers to the representation
\begin{equation}
s(t)=V_0[1+\alpha(t)]\cos[2\pi\nu_{\mu}t+\phi(t)]~.
\end{equation}
The phase noise spectrum $S_\phi(f)$ is of
paramount importance because $\phi$ is related to time, which is the
most precisely measured physical quantity.  For a review on phase
noise see the References \cite{rutman78pieee,ccir90rep580-3,ieee99std1139}.

Most high-speed photodetectors are InGaAs \emph{p-i-n} diodes
operated in strong reverse-bias condition, hence as photoconductors.
Reverse biasing is necessary for high speed because the high electric
field reduces the transit time of the carriers, and also limits the
junction capacitance.  Thus, the depletion region (the intrinsic
layer) can be tailored for quantum efficiency and speed.  The
\emph{p-i-n} diode has the interesting property that even at
relatively low reverse bias $V_b$ ($\sim5$~V) the junction capacitance
is chiefly determined by the thickness of the \emph{i} layer
\cite[pp.\,118--119]{sze:physics-semiconductor-devices},
with little influence from $V_b$.  This indicates that phase noise may
be lower than in other microwave devices.

\section{Experimental method}\label{sec:method}
A preliminary survey of the available detectors shows that none provides
output power sufficient to use a saturated mixer as the phase detector, and
that typical photodetectors have lower noise than common microwave amplifiers.  
Hence we opt for the bridge (interferometric) method, which permits flicker-free amplification before detection.  This method, inspired to
\cite{sann68mtt}, is now a well established technique.  
The full theory and an extensive description of the experimental aspects
is available in \cite{rubiola02rsi}.  Hence, the description given here
focus on the adaptation of the bridge method to the measurement of the 
photodiodes. 

\begin{figure}[t]
\centering
\begin{sideways}
\begin{minipage}{0.81\textheight}
\centering\includegraphics[scale=0.8]{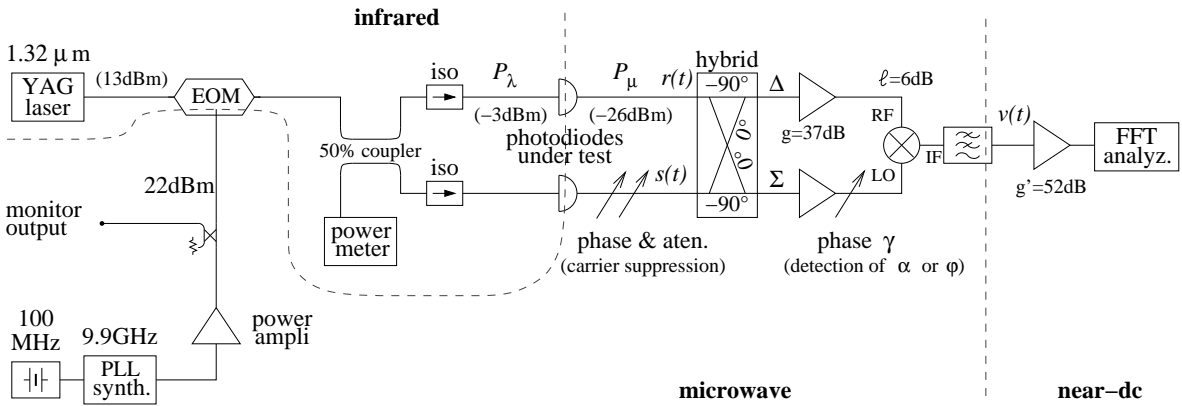}
\vspace*{2em}
\caption{Scheme of the measurement system.}
\label{fig:pdn-scheme}
\end{minipage}
\end{sideways}
\end{figure}

In our configuration (Fig.\,\ref{fig:pdn-scheme}) the 
two detector outputs are combined with appropriate phase and amplitude, so that the sum ($\Sigma$) and the difference ($\Delta$) are available at the output of the hybrid junction.  At the equilibrium condition all of the microwave
power goes in $\Sigma$, while only the imbalance signal, i.e., the photodetector noise plus some residual carrier, is present in $\Delta$.
Close-in flicker noise in amplifiers is a parametric effect that 
results from the flicker fluctuation of the dc bias that modulates the microwave 
carrier.  Of course, the microwave output spectrum is white at zero or very low power. Hence the noise sidebands present in $\Delta$ are amplified without adding flicker.
The $\Sigma$ amplifier provides the power 
needed to saturate the LO port of the mixer, for it flickers. Yet it is shown in 
\cite{rubiola03ell} that the close-in flickering of this amplifier
is not detected because there is no carrier power on the 
other side of the mixer.

The detected signal, converted to dc by the mixer, is
\begin{equation} 
v(t)=k_d\cos(\gamma+\psi)\,\alpha(t)-k_d\sin(\gamma+\psi)\,\phi(t)~,
\end{equation}
where 
$\psi$ is the arbitrary phase that results from the circuit layout.  Thus, the detection 
of amplitude or phase noise is selected by setting the value of $\gamma$. 
A fast Fourier transform (FFT) analyzer measures the output spectrum,
$S_\phi(f)$ or $S_\alpha(f)$. 
The gain, defined as $k_d=v/\alpha$ or $k_d=v/\phi$, is 
\begin{equation}
k_d=\sqrt{\frac{gP_\mu R_0}{\ell}} - 
	\left[\begin{array}{c}\!\!\!\text{dissipative}\!\!\!\\[-0.5ex]
	\text{loss}\end{array}\right]~,
\end{equation}
where $g$ is the amplifier gain, $P_\mu$ the microwave power, $R_0=50$\,\ohm\ the characteristic resistance, and $\ell$ the mixer ssb loss.  Under the conditions  of our setup (see below) the gain is 
$43$ \unit{dBV[/rad]}, including the dc preamplifier.  The notation
\unit{[/rad]} means that /rad appears when appropriate.

Calibration involves the assessment of $k_d$ and the adjustment of $\gamma$.  The gain is measured through the carrier power at the diode output, obtained as the power at the mixer RF port when only one
detector is present (no carrier suppression takes place) divided by the 
detector-to-mixer gain.  This measurement relies on a power meter and on a network analyzer.  The detection angle $\gamma$ is first set by inserting
a reference phase modulator in series with one detector, and nulling the 
output by inspection with a lock-in amplifier.  Under this condition the system detect $\alpha$.  After adding a reference 90\degrees\ to $\gamma$, based either on a network analyzer or on the calibration of the phase shifter, the system detects $\phi$.   The phase modulator is subsequently removed to achieve a 
higher sensitivity in the final measurements. 
Removing the modulator is possible and free from errors because the phase relationship at the mixer inputs is rigidly determined by the carrier suppression in $\Delta$, which exhibits the accuracy of a null measurement.  

The background white noise results from thermal and shot noise.  
The thermal noise contribution is
\begin{equation}
S_{\phi\,t}=\frac{2FkT_0}{P_\mu} +
	\left[\begin{array}{c}\!\!\!\text{dissipative}\!\!\!\\[-0.5ex]
	\text{loss}\end{array}\right]~,
\end{equation}
where $F$ is the noise figure of the $\Delta$ amplifier, and $kT_0\simeq4{\times}10^{-21}$ J is the thermal
energy at room temperature. 
This is proved by dividing the voltage spectrum 
$S_v=\smash{\frac2\ell}gFkT_0$ detected when the $\Delta$ amplifier 
is input-terminated, by the square gain $k_d^2$.
The shot noise contribution of each detector is 
\begin{equation}
S_{\phi\,s}=\frac{4q}{\rho m^2\overline{P}_\lambda}~,
\end{equation} 
where 
$q$ is the electron charge, $\rho$ is the detector responsivity, $m$ 
the index of intensity modulation, and $\overline{P}_\lambda$ 
the average optical power.
This is proved by dividing the spectrum density 
$S_i=2q\overline{\imath}=2q\rho\smash{\overline{P}_\lambda}$
of the the output current $i$ by the average square microwave current
$\overline{i^2_\mathrm{ac}}=\rho^2\smash{\overline{P}^2_\lambda}\,\smash{\frac12}m^2$.  The background amplitude and phase white noise
take the same value because they result from additive
random processes, and because the instrument gain $k_d$ is the same.
The residual flicker noise is to be determined experimentally.

The differential delay of the two branches of the bridge is kept small enough (nanoseconds) so that a discriminator effect does not take place.
With this conditions,  the phase noise of the microwave source and of the 
electro-optic modulator (EOM) is rejected.
The amplitude noise of the source is rejected to the same degree of the carrier attenuation in  $\Delta$, as results from the general properties of the 
balanced bridge.  This rejection applies to amplitude noise and to the laser relative intensity noise (RIN)\@.  

The power of the microwave source is set for the maximum modulation
index $m$, which is the Bessel function $J_1(\cdot)$
that results from the sinusoidal response of the EOM\@.  
This choice also provides increased rejection of the amplitude noise of the microwave source.  
The sinusoidal response of the EOM results in harmonic distortion, 
mainly of odd order; however, these harmonics are out of the system bandwidth.
The photodetectors are operated with some 0.5 mW input power, which
is low enough for the detectors to operate in a linear regime.  
This makes possible a high carrier suppression (50--60 dB) in $\Delta$, which is stable for the duration of the measurement (half an hour), and also provides a high rejection of the laser RIN and of the noise of the $\Delta$ amplifier.
The coherence length of the YAG laser used in our experiment is about
1 km, and all optical signals in the system are highly coherent.

\section{Results}\label{sec:results}
\begin{figure}[t]
\centering\includegraphics[bb=62 193 528 510,clip,scale=0.6]{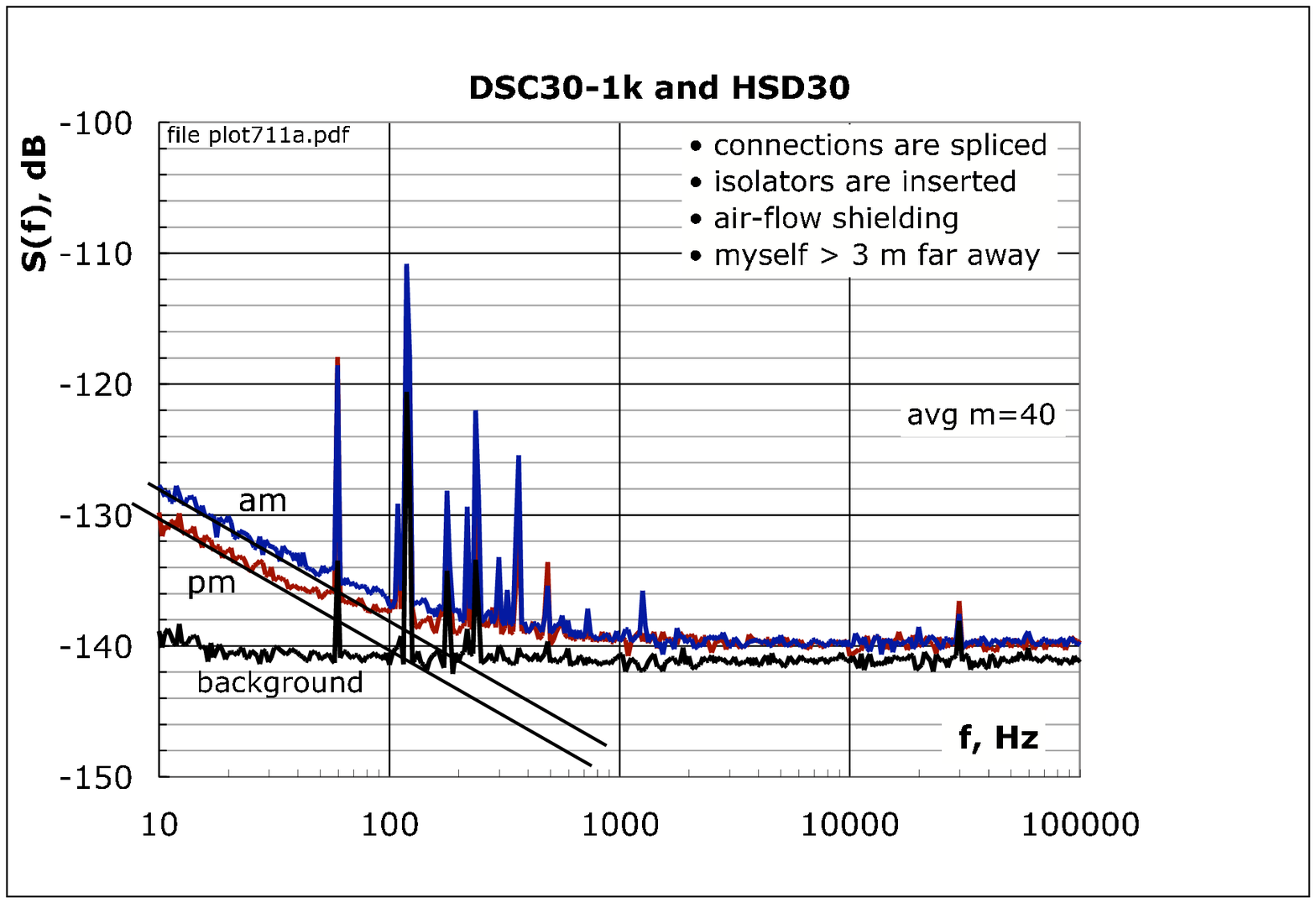}
\caption{Example of measured spectra $S_\alpha(f)$ and $S_\phi(f)$.}
\label{fig:pdn-plot711}
\end{figure}
The background noise of the instrument is measured in two steps.  
A first value is measured by replacing the photodetectorsÕ output with 
two microwave signals of the same power, derived from the main source.
The noise of the source is rejected by the bridge measurement.
A more subtle mechanism, which is not detected by the first measurement,
is due to the fluctuation of the mixer offset voltage induced by the
fluctuation of the LO power~\cite{brendel77im}.
This effect is measured in a second test, by restoring the photodetectors and 
breaking the path from the hybrid junction to the $\Delta$ amplifier, and terminating the two free ends.
The worst case is used as the background noise.  
The background thereby obtained places an upper bound for the
$1/f$ noise, yet hides the shot noise. This is correct because the shot 
noise arises in the photodiodes, not in the instrument. 
The design criteria of Sec.\ \ref{sec:method} result in a 
background flicker of approximately $-135$ \unit{dB[rad^2]/Hz} 
at $f=1$ Hz, hardly visible above 10 Hz (Fig.\ \ref{fig:pdn-plot711}).
The white noise, about $-140$ \unit{dB[rad^2]/Hz}, is close to the
expected value, within a fraction of a decibel.  It is used only as a diagnostic check, to validate the calibration.

We tested three photodetectors, a Fermionics HSD30, a Discovery
Semiconductors DSC30-1k, and a Lasertron QDMH3.  These devices
are InGaAs \emph{p-i-n} photodiodes suitable to the wavelength of
1.3 $\mu$m  and 1.55 $\mu$m, exhibiting and a bandwidth in excess 
of 12 GHz, and similar to one another.  They are routinely used in our photonic oscillators~\cite{yao96josab,yao97ol} and in related experiments.  

Each measurement was repeated numerous times with different averaging samples in order to detect any degradation from low-frequency or non-stationary phenomena, if present.
Figure~\ref{fig:pdn-plot711} shows an example of the measured spectra.
Combining the experimental data, we calculate the flicker of each device,  shown in  Table~\ref{fig:pdn-single-detector}.
Each spectrum is affected by a 
random uncertainty is of 0.5 dB, due to the parametric spectral estimation
(Ref.~\cite{percival:spectrum-analysis},~chap.\,9),
and to the measurement of the photodetector output power.
In addition, we account for a systematic uncertainty of  1~dB
due to the calibration of the gain.  
The random uncertainty is amplified in the process
of calculating the noise of the individual detector from the available spectra.  Conversely, the systematic uncertainty is a constant error that applies to all measurements, for it is not amplified.

\begin{table}[t]
\caption{Flicker noise of the photodiodes.}
\label{fig:pdn-single-detector}
\begin{center}
\begin{tabular}{lcccc}\hline
\rule[0ex]{0ex}{2.5ex}%
photodiode  & \multicolumn{2}{c}{$S_\alpha(1\unit{Hz})$}
                  & \multicolumn{2}{c}{$S_\phi(1\unit{Hz})$}\\
\rule[-1ex]{0ex}{0ex}& estimate & uncertainty & estimate & uncertainty \\
\hline
\rule[-2ex]{0ex}{5.5ex}%
HSD30    & $-122.7$& {\small\begin{tabular}{r}%
                      $-7.1$\\[-0.5ex]$+3.4$\end{tabular}}
         & $-127.6$& {\small\begin{tabular}{r}%
                      $-8.6$\\[-0.5ex]$+3.6$\end{tabular}}\\  \hline
\rule[-2ex]{0ex}{5.5ex}%
DSC30-1K & $-119.8$& {\small\begin{tabular}{r}%
                      $-3.1$\\[-0.5ex]$+2.4$\end{tabular}}
         & $-120.8$& {\small\begin{tabular}{r}%
                      $-1.8$\\[-0.5ex]$+1.7$\end{tabular}}\\  \hline
\rule[-2ex]{0ex}{5.5ex}%
QDMH3    & $-114.3$& {\small\begin{tabular}{r}%
                      $-1.5$\\[-0.5ex]$+1.4$\end{tabular}}
         & $-120.2$& {\small\begin{tabular}{r}%
                      $-1.7$\\[-0.5ex]$+1.6$\end{tabular}}\\\hline
\rule[0ex]{0ex}{2.5ex}%
unit & dB/Hz & dB & \unit{dBrad^2/Hz} & dB
\end{tabular}
\end{center}
\end{table}

\section{Discussion}\label{sec:discussion}
\begin{figure*}[t]
\def\thisfigurescalefactor{1.4}
\centering%
\scalebox{\thisfigurescalefactor}{%
\includegraphics[bb=75 170 536 530,clip,scale=0.26]{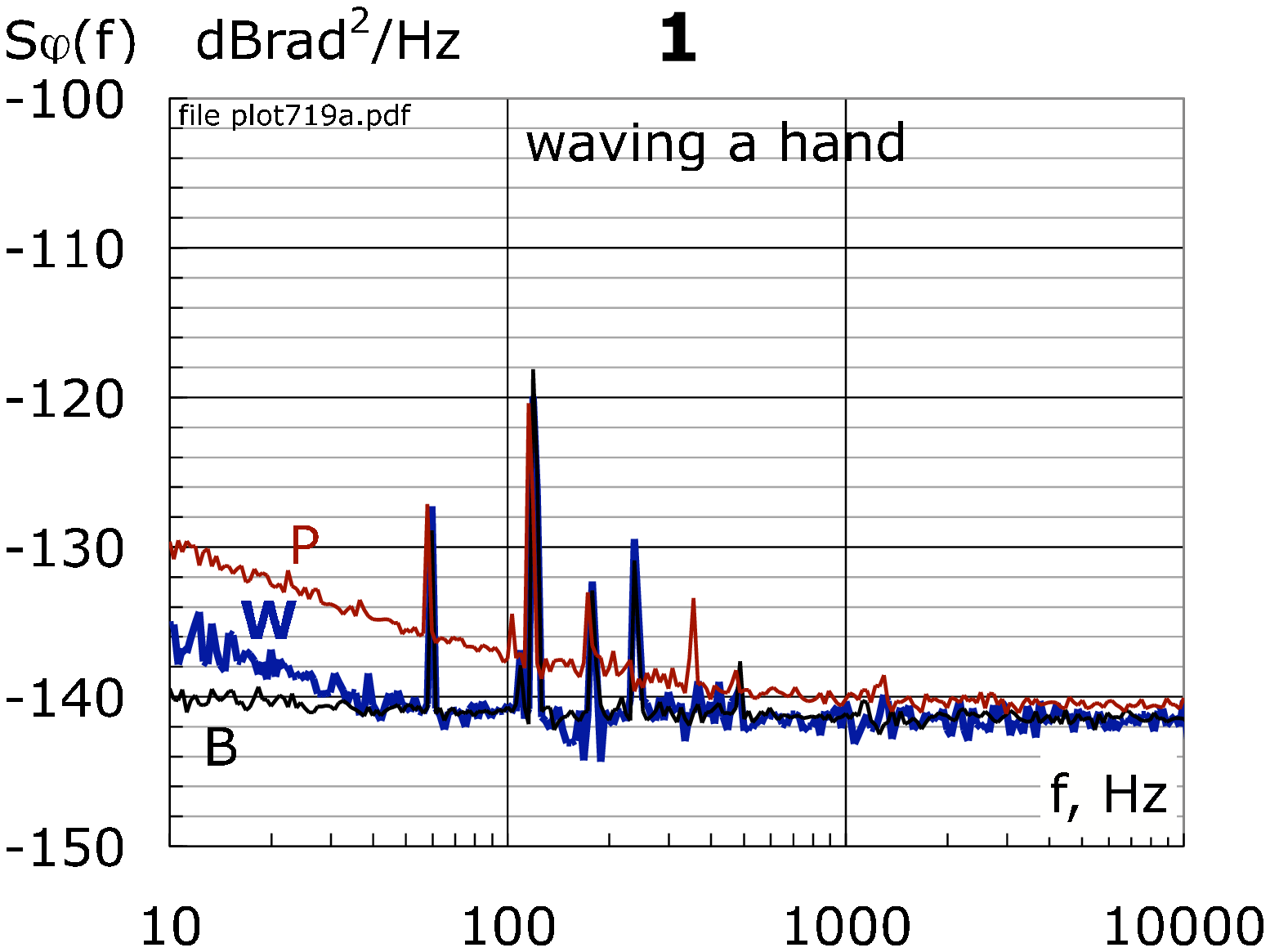}
\includegraphics[bb=75 170 536 530,clip,scale=0.26]{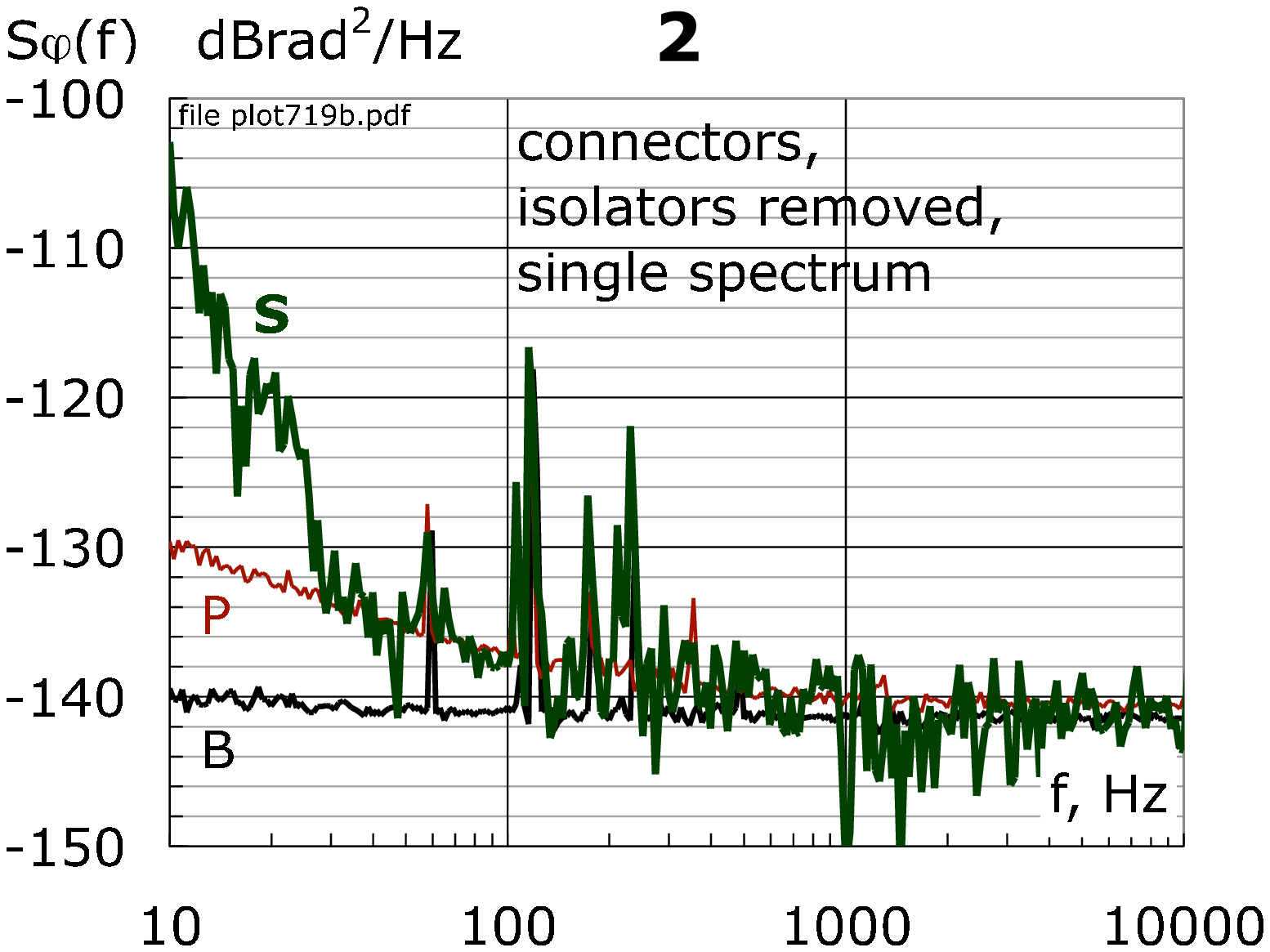}}\\
\scalebox{\thisfigurescalefactor}{%
\includegraphics[bb=75 170 536 530,clip,scale=0.26]{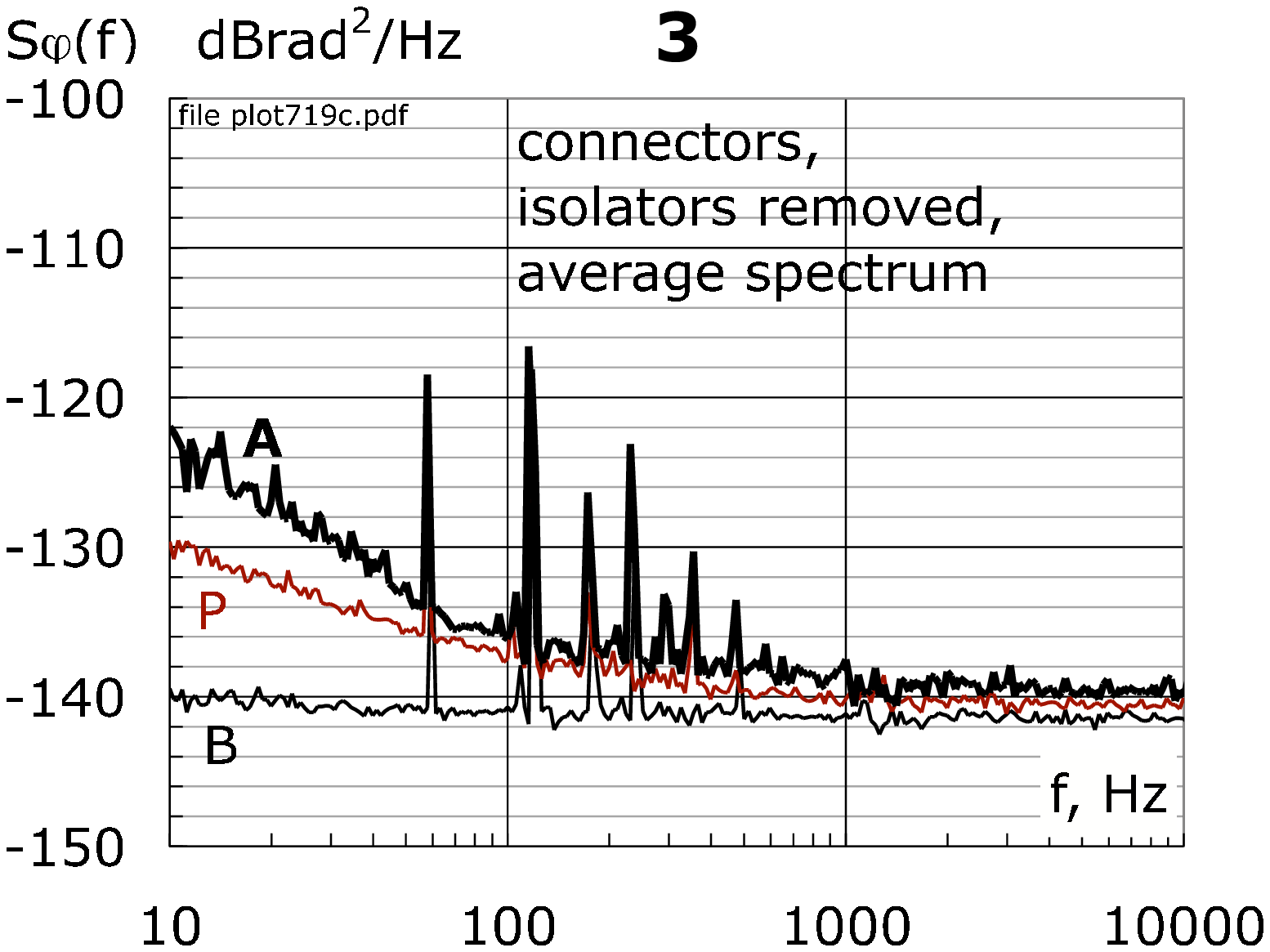}
\includegraphics[bb=75 170 536 530,clip,scale=0.26]{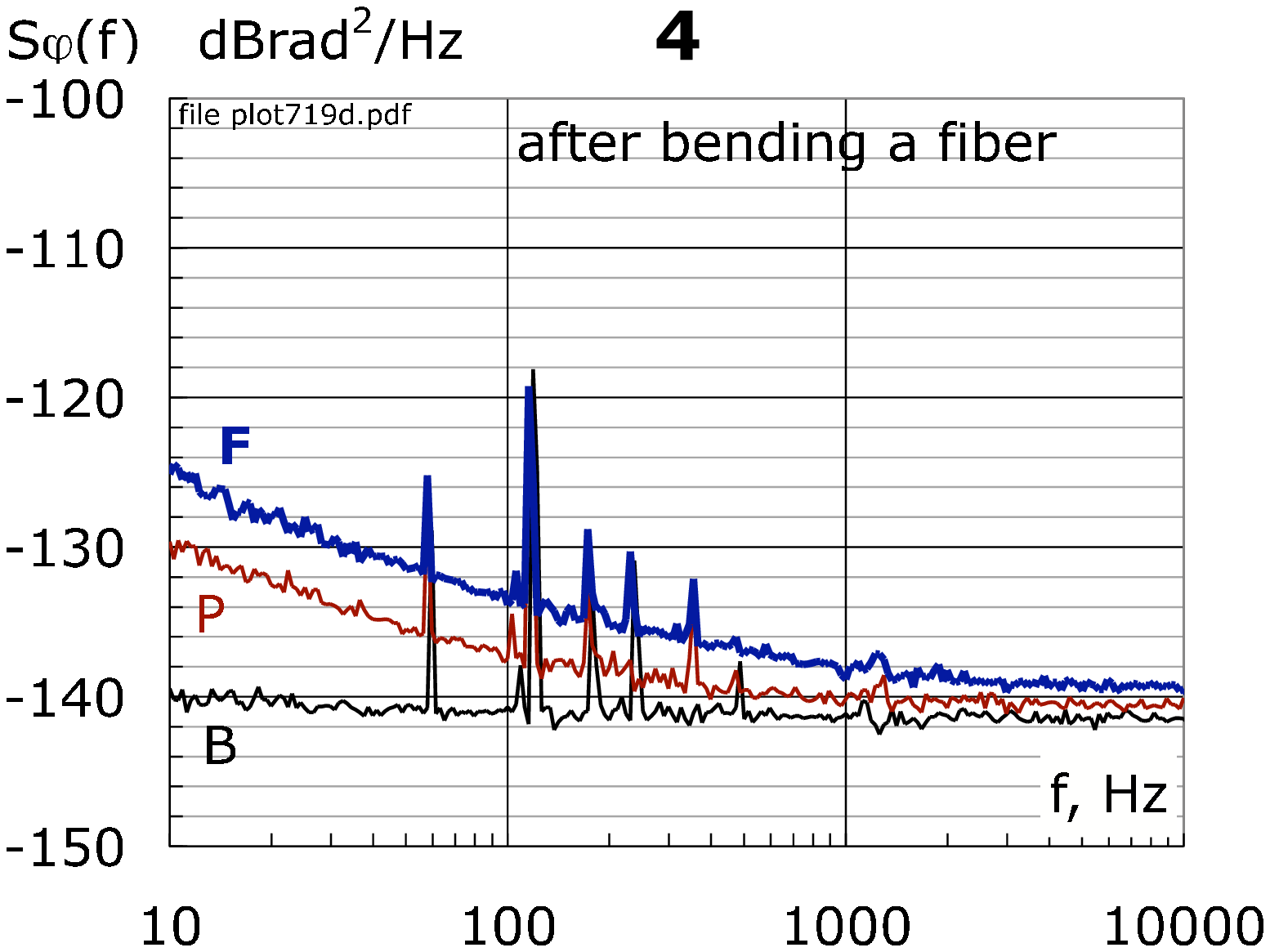}}
\caption{Examples of environment effects and experimental mistakes around the corner.
All the plots show the instrument \textsf{B}ackground noise (spectrum \textsf{B}) and the noise spectrum of the \textsf{P}hotodiode pair (spectrum \textsf{P}).
Plot \textsf{1} spectrum \textsf{W}: the experimentalist \textsf{W}aves a hand gently ($\approx0.2$ m/s), 3 m far away from the system.  
Plot \textsf{2} spectrum \textsf{S}: the optical isolators are removed and the connectors are restored at the input of the photodiodes (\textsf{S}ingle spectrum).
Plot \textsf{3} spectrum \textsf{A}: same as plot 3, but \textsf{A}verage spectrum.
Plot \textsf{4} spectrum \textsf{F}: a \textsf{F}iber is bended with a radius of $\approx5$ cm, which is twice that of a standard reel.}  
\label{fig:pdn-plot719}
\end{figure*}

For practical reasons, we selected the configurations that give
reproducible spectra with low and smooth $1/f$ noise that are not influenced by the sample averaging size.  Reproducibility is related
to smoothness because technical noise shows up at very low frequencies,
while we expect from semiconductors smooth $1/f$ noise in a wide frequency
range.  Smoothness was verified by comparison with a database
of trusted spectra.  
Technical noise turned out to be a serious difficulty.  As no data 
was found in the literature, we give some practical hints in Fig.\ \ref{fig:pdn-plot719}.

The EOM requires a high microwave power  (20 dBm or more), which is some 50 dB higher than the photodetector output.  The isolation in the microwave circuits is hardly higher than about 120 dB\@.
Thus crosstalk, influenced by the fluctuating dielectric constant of the
environment, turns into a detectable signal.
The system clearly senses the experimentalist waving a hand 
($\approx0.2$ m/s) at a distance of 3 m.  The spectrum (Fig.\ \ref{fig:pdn-plot719}.1, plot W) is easily taken for flicker.
This problem can be mitigated using the new high-efficiency EOMs \cite{vaneck02:polymer-modulators}.

Air flow affects the delay of the optical fibers, thus some isolation
is necessary to mitigate this effect.
All our attempts failed until we inserted optical isolators in
series with the photodetectors, and spliced all the fiber junctions
(except the laser output).  After this, the back-reflected light
at the unused port of the coupler was below the sensitivity
of the power-meter, which is 1~nW\@.
Without isolation and splicing, individual spectra show
spikes appearing at random times (Fig.\ \ref{fig:pdn-plot719}.2, plot S).
Averaging yields a smooth spectrum.  Yet slope is incorrect (Fig.\ \ref{fig:pdn-plot719}.3, plot A).
Beside the mechanics of the connectors, we attribute
this effect to reflection noise in the optical fibers \cite{gimlett89jlt,shieh98ptl}.

Even after isolating and splicing, we observed that bending a fiber may result in increased flickering.  Afterwards, the spectrum may become irregular, or still be smooth with a clean $1/f$ slope, as in Fig.\ \ref{fig:pdn-plot719}.4, plot F,
but nevertheless incorrect. 
We interpret this as a change in the interference pattern in the fiber
due to polarization.
The observed increase in noise is clearly systematic, although reproducing the numerical value takes some effort.

Spectral lines at 60 Hz and its multiples are present in the noise spectra, originated by magnetic fields, in all cases lower than $-110$ \unit{dB[rad^2]/Hz}.
The level of these stray signals is about the same found routinely
in the phase noise measurement with the saturated mixer method, yet with a carrier power of some 10 dBm instead of the $-26$ dBm of our experiments, thus with a signal-to-noise ratio proportionally higher.   
The superior immunity of the bridge scheme is due to microwave amplification of the noise sidebands before detecting.

The $1/f$ spectra of the detectors we measured are similar, and a value of $-120$~\unit{dB[rad^2]/Hz} at $f=1$~Hz can be taken as representative of both amplitude and phase noise.  Using the formulae available in 
\cite{rutman78pieee,ccir90rep580-3,ieee99std1139}, a spectrum of the form $h_{-1}/f$ converted into the Allan (two-sample) variance $\sigma^2(\tau)$ is $\sigma^2=2\ln(2)\:h_{-1}$ independent of the measurement time $\tau$.  
The length of 1 rad ®in a fiber of refraction index $n=1.45$, at the modulation frequency $\nu_\mu=9.9$ GHz, of is 3.3 mm. 
Thus a phase noise of $-120$~\unit{dBrad^2/Hz} at $f=1$~Hz ($h_{-1}=10^{-12}$) is equivalent to a fluctuation $\sigma_l(\tau)=3.9$ nm of the optical length $l$.

\section{Final remarks}\label{sec:conclusions}
It is generally accepted \cite{sikula:icnf03} that flicker noise is an elusive phenomenon and that our understanding is based on models, the most
accreditated of which are due to Hooge \cite{hooge69pla} and to Mc{W}horter
\cite{mcwhorter57}, rather than on a unified theory.
On the other hand, the presence of the phase and amplitude flickering in a microwave carrier is believed to be the dc flicker, up-converted by a nonlinearity.
This also applies to the photodiode, even though in this case the dc bias exists only in the presence of light.  In fact, removing the modulation results in a white microwave spectrum, flat around any frequency in the passband of the ststem.

The experimental difficulties we encountered are due to various forms of technical noise, at an exceedingly low level, which nevertheless may exceed the detector noise, unless great care is taken.  
On one hand, this means that the environment in which the diode is inserted must be revisited if one needs the lowest achievable noise. On the other hand, this means that the photodiode exhibits low noise and high 
stability, and that it has an unexploited potential for new and emerging applications.

\section*{Acknowledgements}
The research described in this paper was carried out at the Jet
Propulsion Laboratory, California Institute of Technology, under
contract of the National Aeronautics and Space Administration, and
with support from ARL and AOSP / DARPA\@.  We thank the
Universit\'e Henri Poincar\'e for partially supporting E. Rubiola while 
visiting JPL, and F. Lardet-Vieudrin of FEMTO-ST, France, for providing low-flicker dc
preamplifiers.

\def\bibfile#1{/Users/rubiola/Documents/work/bib/#1}
\bibliographystyle{amsalpha}
\bibliography{\bibfile{ref-short},%
              \bibfile{references},%
              \bibfile{rubiola}}

\end{document}